\documentclass[times,twocolumn,final]{elsarticle}
\biboptions{sort&compress}
\usepackage{cnf}

\usepackage{framed,multirow}

\usepackage{amssymb}
\usepackage{latexsym}

\usepackage{url}
\usepackage{xcolor}
\definecolor{newcolor}{rgb}{.8,.349,.1}
\usepackage{amsmath}
\usepackage{units}
\usepackage{mathcomp}
\usepackage{mathrsfs}
\usepackage{hyperref}
\usepackage{soul}
\usepackage[switch,pagewise]{lineno} 
\journal{Combustion and Flame}

\begin{document}

\verso{Rostami et al.}

\begin{frontmatter}

\title{Experimental investigation of the turbulent kinetic energy budget in small Lewis number premixed flames}

\author[]{Ali \snm{Rostami}\corref{cor1}}
\author[]{Sina \snm{Kheirkhah}}
\cortext[cor1]{Corresponding author: School of Engineering, The University of British Columbia, Kelowna, Canada, V1V 1V7}
\emailauthor{ali.rostami@ubc.ca}{Ali Rostami}

\address{School of Engineering, The University of British Columbia, Kelowna, Canada, V1V 1V7}

\begin{abstract}
The Conditioned Particle Image Velocimetry technique is used to investigate the effects of flame on the turbulent kinetic energy and its transport within the flame brush of small-Lewis-number premixed flames. Turbulent hydrogen-enriched methane-air flames, with the volume fraction of hydrogen ranging from 0 to 80\%, are stabilized using a Bunsen burner. The turbulent characteristics of the background flow and the unstretched laminar flame speed of the examined flames are fixed. The effective Lewis number and the Karlovitz number of the studied flames vary from 1.03 to 0.71 and from 0.84 to 0.96, respectively. The results show that, for flames with an effective Lewis number less than unity, the turbulent kinetic energy is maximized within the flame brush. Specifically, for the examined flames with the smallest effective Lewis number, the maximum turbulent kinetic energy within the flame brush is approximately five times that in the non-reacting flow, indicating a substantial augmentation of the turbulent kinetic energy within the flame brush. The assessment of several terms in the turbulent kinetic energy budget shows that while the term relevant to the mean pressure gradient serves as a main source of the turbulent kinetic energy, the dilatation term is the dominant sink among the examined terms. The turbulent flux of the turbulent kinetic energy features the counter-gradient behavior, which is enhanced with decreasing the effective Lewis number. While these observations are consistent with the findings of past Direct Numerical Simulations of statistically-planar turbulent premixed flames, our results extend past findings to experimentally examined laboratory-scale Bunsen flames.

\end{abstract}


\end{frontmatter}


\section*{Novelty and significance statement}

The effect of combustion on modifying turbulence within the flame brush of small-Lewis-number jet flames has not been investigated. Our findings show that turbulent kinetic energy is substantially augmented in turbulent jet flames with Lewis numbers smaller than unity. Here, we experimentally show for the first time that the combustion-induced dilatation contributes to the dissipation of the turbulent kinetic energy, while the flame-induced mean pressure gradient contributes to the production of the turbulent kinetic energy in small-Lewis-number flames. Our findings point to the occurrence of counter-gradient turbulence transport that standard gradient-based models would fail to capture. The above findings are of significance for fundamental understanding of flame-turbulence interaction and for the assessment and development of combustion closure models.

 
\section{Introduction}

Turbulent premixed combustion is the mode of operation in various practical combustion systems and involves complex flame-turbulence interactions. These interactions are coupled and occur in both ways: turbulence modifying the flame characteristics and the flame changing the turbulent flow characteristics \cite{lipatnikov2010effects}. Flame-turbulence interactions have been the subject of research among the combustion community, and many articles have reviewed this topic, see for example \cite{lipatnikov2010effects,driscoll2008turbulent, chakraborty2021influence,steinberg2021structure}. Past studies have investigated the influences of flame on turbulence in terms of the fluctuating velocity field \cite{driscoll1988measurement}, Reynolds stresses \cite{lee2020heat}, enstrophy  \cite{lipatnikov2014direct,kazbekov2021flame,bobbitt2016vorticity,rising2021effects,geikie2021turbulent}, and turbulent kinetic energy \cite{ driscoll1988measurement, bray1981turbulence,nishiki2002modeling, chakraborty2011effects}. The review of literature shows that only a few Direct Numerical Simulation (DNS) studies \cite{chakraborty2011effects, su2022suppression} have examined the influences of flame on the turbulent kinetic energy budget in small effective Lewis number ($Le$) premixed flames. Despite the importance of the DNS studies, to our best knowledge, no experimental counterpart exists and this motivates the present investigation.

The influence of flame on turbulence was first examined in early experimental studies of~\cite{karlovitz1953studies,scurlock1953propagation}. There~\cite{karlovitz1953studies,scurlock1953propagation}, it was observed that the measured turbulent burning velocity was larger than that estimated using the Damk\"{o}hler model, which was developed using the reactant's Root-Mean-Square (RMS) velocity fluctuations~\cite{lipatnikov2010effects}. Elaborating the above observation, it was hypothesized in \cite{karlovitz1953studies} that the thermal expansion due to the heat release rate leads to preferential acceleration of combustion products compared to the fresh reactants in the direction normal to the instantaneous flame front. Such preferential acceleration results in flame-induced velocity fluctuations (also referred to as the flame-generated turbulence) in addition to those induced by turbulent eddies of cold flow~\cite{chakraborty2021influence}. The enhanced velocity fluctuations were discussed to increase the turbulent burning velocity. The hypothesis in \cite{karlovitz1953studies} was supported later using the experimental studies in which the statistics of the conditioned velocity were measured, and it was confirmed that, in weakly turbulent flames, the velocity fluctuations of combustion products could be indeed larger than those of the reactants~\cite{driscoll1988measurement, cheng1984conditional,pfadler2008systematic}. 

Assessing the influence of flame on turbulence, the turbulent kinetic energy can be estimated within the flame brush. Care should be taken to differentiate between ``true'' velocity fluctuations (caused by turbulent eddies) versus fluctuations caused by the intermittent movement of the flame front at a given measurement location~\cite{driscoll1988measurement}. For a bimodal distribution of the combustion progress variable ($c$), i.e., the thin flame assumption, Bray~\textit{et al}. \cite{bray1985unified} showed that the Favre-averaged turbulent kinetic energy ($\tilde{k}$) equals the summation of the true turbulent kinetic energy ($k_\mathrm{t}$) and that due to the unburned-burned intermittency ($k_\mathrm{int}$). That is 

\begin{equation}
\begin{split}
\label{Eq:Bray}
\tilde{k}
= \frac{1}{2}\widetilde{u_i''^{\,2}}
={}&
\underbrace{
(1-\tilde{c})
\underbrace{
\frac{1}{2}
\overline{
\left(
u_{i,\mathrm{r}}-\overline{u_{i,\mathrm{r}}}
\right)^2
}
}_{k_{\mathrm{r}}}
+
\tilde{c}
\underbrace{
\frac{1}{2}
\overline{
\left(
u_{i,\mathrm{p}}-\overline{u_{i,\mathrm{p}}}
\right)^2
}
}_{k_{\mathrm{p}}}
}_{k_{\mathrm{t}}}
\\
&+
\underbrace{
\frac{1}{2}
\tilde{c}(1-\tilde{c})
\left(
\overline{u_{i,\mathrm{p}}}
-
\overline{u_{i,\mathrm{r}}}
\right)^2
}_{k_{\mathrm{int}}}.
\end{split}
\end{equation}
In Eq.~(\ref{Eq:Bray}), $u_{i}$ is the velocity along the $i-$direction, and the overbar and tilde operations denote the Reynolds and Favre averaging, respectively. In Eq.~(\ref{Eq:Bray}), $\widetilde{u_i''}=u_{i}-\widetilde{u_{i}}$, and the indices p and r refer to the products and reactants, respectively. As can be seen in Eq.~(\ref{Eq:Bray}), $k_\mathrm{t}$ is the weighted-summation of the turbulent kinetic energy of the reactants ($k_\mathrm{r}$) and products ($k_\mathrm{p}$). The intermittent turbulent kinetic energy can be non-zero even when the flow in the reactants and products is laminar (i.e., $k_\mathrm{p}=k_\mathrm{r}=0$), but the flame moves back and forth leading to $\tilde{k} \neq 0$~\cite{lipatnikov2010effects}. Driscoll and Gulati \cite{driscoll1988measurement} experimentally investigated the effects of the heat release parameter ($\tau = T_\mathrm{p}/T_\mathrm{r}-1$, with $T_\mathrm{p}$ and $T_\mathrm{r}$ being the product and reactant temperatures, respectively) on $\tilde{k}$ for turbulent premixed methane/air flames and their results showed that increasing $\tau$ increases $\tilde{k}$ in the flame brush.

The mechanisms contributing to suppression or augmentation of the turbulent kinetic energy within the flame brush are determined by the analysis of the transport equation of $\tilde{k}$, which is given by \cite{lipatnikov2010effects} 

\begin{equation}
\begin{split}
\label{Eq.TKETransport}
\frac{\partial (\overline{\rho} \tilde{k})}{\partial t}
+\underbrace{\frac{\partial}{\partial x_i}
\left(\overline{\rho}\,\tilde{u}_i \tilde{k} \right)}_{T_1}=\underbrace{-\overline{\rho\,u_j''u_i''}\frac{\partial \tilde{u}_j}{\partial x_i}}_{T_2}\underbrace{-\overline{u_i''}\frac{\partial \overline{p}}{\partial x_i}}_{T_3} & \\
\underbrace{-\frac{\partial}{\partial x_i}\overline{\rho\,u_i''k}}_{T_4}\underbrace{-\overline{u_i''\frac{\partial p'}{\partial x_i}}}_{T_5}\underbrace{+\overline{u_j''\frac{\partial \mathcal{T}_{ji}}{\partial x_i}}}_{T_6},
\end{split}
\end{equation}
where $t$, $\rho$, $p$, and $\mathcal{T}_{ji}$ are time, density, pressure, and viscous stress tensor, respectively. In Eq.~(\ref{Eq.TKETransport}), $T_{1}$ contributes to the transport of $\tilde{k}$ via convection and is referred to as the convection term. $T_{2}$ and $T_{3}$ contribute to the turbulent kinetic energy transport via the gradient of the mean velocity and the mean pressure, respectively. The former and the latter are referred to as the mean dilatation and pressure work terms, respectively. $T_{4}$ and $T_{5}$ contribute to the transport of $\tilde{k}$ via velocity and pressure fluctuations, respectively. $T_{6}$ is referred to as the molecular viscous dissipation term. 

A few DNS studies~\cite{ chakraborty2011effects, macart2018effects,su2022suppression} have reported the variation of $T_{1-6}$ across the flame brush of turbulent premixed flames and examined the effects of varying $Le$ and the Karlovitz number ($Ka$, which is the ratio of chemical to turbulence time scales) on $T_{1-6}$. For statistically-planar turbulent premixed flames with $Ka=13$, Chakraborty \textit{et al.} \cite{chakraborty2011effects} studied the effects of decreasing $Le$ (for $Le\leq 1.2$) on $T_{1-6}$. Their results~\cite{chakraborty2011effects} showed that for flames with $Le\gtrsim 0.8$, the contributions of $T_{2-5}$ to the budget of $\tilde{k}$ were relatively small, and the dissipation term ($T_6$) was dominant and negative, leading to net dissipation of the turbulent kinetic energy across the flame brush of flames with $Le\gtrsim0.8$. For flames with $Le \lesssim 0.6$, while $T_6$ was negative, $T_3$ was relatively large and positive, leading to the net production of $\tilde{k}$ across the flame brush. For turbulent premixed jet flames, in which the shear layer generates velocity gradient in the direction transverse to the flow main direction, Su \textit{et al.} \cite{su2022suppression} studied the effects of decreasing $Le$ (for $Le\leq 1.6$) on the budget of $\tilde{k}$. Their analysis~\cite{su2022suppression} showed that while the major sink term remained $T_6$ for their examined flames, the primary source term changed from $T_2$ to $T_3$ with decreasing $Le$. MacArt \textit{ et al.}~\cite{macart2018effects} examined the effects of increasing $Ka$ on $T_{1-6}$ for turbulent premixed jet flames. Their results~\cite{macart2018effects} showed that for flames with Karlovitz number smaller (larger) than a critical value, given by $Ka_\mathrm{cr}=\rho_{\mathrm{r}}/\rho_{\mathrm{p}}-1$, with $\rho_{\mathrm{p}}$ and $\rho_{\mathrm{r}}$ being the combustion product and reactant densities, $T_2$ is negative (positive), meaning that $T_2$ contributes to dissipation (production) of $\tilde{k}$ in flames with $Ka<Ka_\mathrm{cr}$ ($Ka>Ka_\mathrm{cr}$). Their analysis \cite{macart2018effects} showed that for flames with $Ka<Ka_\mathrm{cr}$, $T_2$ is the dominant sink term, while $T_3$ and $T_5$ are the dominant source terms with $T_3+T_5>|T_2|$, leading to the net production of $\tilde{k}$ within the flame brush. However, for flames with $Ka>Ka_\mathrm{cr}$, while $T_3$ and $T_5$ were negligible, $T_2$($T_6$) was the main source (sink) term. In summary, at large $Ka$, the budget of $\tilde{k}$ is similar to its non-reacting counterpart, in which turbulence is generated due to shear flow and is dissipated by viscous dissipation~\cite{macart2018effects}. To our best knowledge, only one study \cite{driscoll1988measurement} has examined the constituent terms of Eq.~(\ref{Eq.TKETransport}) experimentally. In Ref.~\cite{driscoll1988measurement}, the Laser Doppler Velocimetry and the Rayleigh scattering technique were used to measure $T_{1}$, $T_{2}$, $T_{3}$, and $T_{4}$ in unconfined, oblique, methane/air premixed flames. The results in~\cite{driscoll1988measurement} showed that $T_{3}$ ($T_{2}$) is the major source (sink) of $\tilde{k}$. 

Although past DNS studies \cite{chakraborty2011effects, su2022suppression} have reported the effects of $Le$ on $\tilde{k}$ and its transport, no investigation has experimentally examined these effects for laboratory-scale flames. The first objective of the present study is to investigate the effects of decreasing $Le$ on $\tilde{k}$ in hydrogen-enriched methane-air turbulent premixed flames with $Le \leq 1.03$. The second objective is to measure $T_{1-4}$ in Eq.~(\ref{Eq.TKETransport}) and understand how decreasing $Le$ affects the transport of $\tilde{k}$. It is acknowledged that measuring $T_{5}$ and $T_{6}$ was not possible in the present study due to limitations of the utilized diagnostics, and as such, the discussions are presented considering our experimental limitation.

\section{Experimental methodology}

\begin{figure}[!h]
	\centerline{\includegraphics[width=0.43\textwidth]{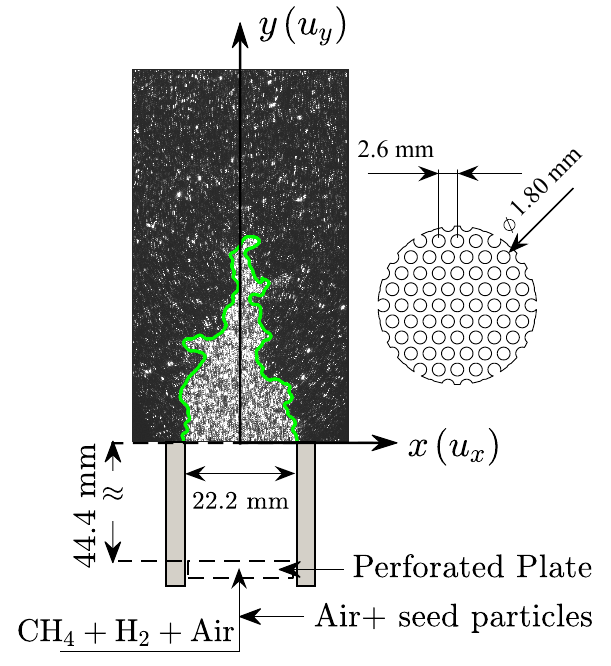}}
	\caption{A cropped raw Mie-scattering image of turbulent premixed flame with $Le=0.71$ along with the technical drawing of the perforated plate. The overlaid green curve on the Mie scattering image is the detected flame front.} 
	\label{Fig:Setup}
       \vspace{-6 mm}
\end{figure}

A Bunsen burner, with an exit diameter of \unit[22.2]{mm}, was used in the present study. The burner is identical to that in~\cite{mohammadnejad2020thick}, and further details regarding the technical drawing of the burner can be found in there. The background turbulent flow was produced using a perforated plate which includes \unit[1.8]{mm} circular holes arranged
in a hexagonal pattern (see the technical drawing in Fig.~\ref{Fig:Setup}). This  plate has a blockage ratio of
60\%  and was positioned \unit[44.4]{mm} upstream of the burner exit plane (see the schematic drawing in Fig.~\ref{Fig:Setup}). The turbulent premixed flames were stabilized using a diffusion hydrogen pilot flame with a small flow rate (less than 2\% of the reactants flow rate).

The Conditional Particle Image Velocimetry (CPIV) was the primary diagnostic of the present study and was used to measure several terms of Eqs.~(\ref{Eq:Bray}) and (\ref{Eq.TKETransport}). In the CPIV technique, the flame front is detected using the significant change in the intensity of PIV seed particles (see a representative detected flame front in Fig.~\ref{Fig:Setup} shown by the green curve), and is used to condition the velocity field into the reactant and product regions~\cite{pfadler2008systematic, rostami2026turbulent}. The hardware of the CPIV  techniques was identical to that in ~\cite{rostami2024astigmatic,rostami2026turbulent}. The CPIV hardware included a dual-cavity Nd:YAG laser (wavelength of \unit[532]{nm}), sheet-forming optics, a programmable timing unit, and two sCMOS cameras equipped with Scheimpflug adapters and collection optics. The latter consisted of two \unit[85]{mm}-focal-length lenses equipped with \unit[532$\pm$10]{nm} bandpass filters to suppress the flame luminosity in the collected PIV images. The cameras were placed on the same side of the CPIV laser sheet, and the angle between the viewing directions of these cameras was about $30^{\circ}$. The thickness of the collimated laser sheet at the measurement plane was about \unit[0.5]{mm}. The flow was seeded with sub-micron Aerosil\textsuperscript{\textregistered} particles. Assuming an agglomerated size of \unit[0.5]{$\tcmu$m}, a particle density of $\unit[50]{\mathrm{kg\,m^{-3}}}$ (provided in the data sheet of the seed particle manufacturer), and considering the smallest length scale of the non-reacting flow (to be discussed later), the seed particle's Stokes number was about 0.00003, confirming the suitability of the utilized seed particles for CPIV. The Field-Of-View (FOV) of the dewarped CPIV images was \unit[80]{mm} (horizontal) by \unit[71]{mm} (vertical), which corresponded to a pixel resolution of \unit[35]{$\tcmu$m}. For each test condition, 1000 image pairs were acquired with an acquisition frequency of \unit[10]{Hz}. The LaVision DaVis 10.2.1 software was used to calculate the velocity field using a spatial cross-correlation algorithm. The interrogation window size was set to $32\times32$~pixels$^{2}$ with an overlap of $50\%$ between the windows, which resulted in a vector spacing of \unit[0.56]{mm}. To ensure that the seed particles' maximum displacement was restricted to 25\%~\cite{picano2011dynamics} of the interrogation window size ($8$~pixels), the two laser pulses were separated by \unit[47]{$\tcmu$s}. The average uncertainty in measuring the instantaneous in-plane velocity was quantified using the correlation statistics approach \cite{wieneke2015piv} and was about \unit[0.15]{$\mathrm{m\,s^{-1}}$}. 

A Cartesian coordinate system, with its origin positioned at the intersection of the burner centerline ($y-$axis) and the exit plane of the burner, was used (see Fig.~\ref{Fig:Setup}). The $x-$axis coincided with the direction of the CPIV laser sheet, and the $z-$axis was perpendicular to both $x$ and $y-$axes. The velocities along $x$, $y$ and $z$-directions are denoted by $u_x$, $u_y$, and $u_z$, respectively.

Five operating conditions (Flames A--E) were tested, with details presented in Table~\ref{tab: Test conditions}. The fuel was a mixture of hydrogen and methane, and the volume fraction of hydrogen in the fuel mixture ($x_\mathrm{H_2}$) varied from 0 to 80\%. To isolate the influence of the cold-flow turbulence on flame characteristics, the mean bulk flow velocity was kept constant at \unit[5.0]{$\mathrm{m\,s^{-1}}$}. The unstretched laminar flame speed ($S_\mathrm{L,0}$) was fixed at \unit[0.195]{$\mathrm{m\,s^{-1}}$}. This parameter was obtained from Cantera~\cite{cantera} simulations of a freely propagating laminar flame with GRI-Mech 3.0~\cite{GRI_Mech}. To assess the sensitivity of $S_\mathrm{L,0}$ calculation to the chemical reaction mechanism, the Cantera simulations were also performed using the FFCM-2~\cite{ZDV2023} mechanism. Using this mechanism, a maximum deviation of approximately 6\% in calculating $S_\mathrm{L,0}$ was obtained compared to that from GRI-Mech 3.0. To keep $S_\mathrm{L,0}$ constant while changing $x_\mathrm{H_2}$, the fuel-air equivalence ratio ($\phi$) varied across the examined flames. The definition provided in \cite{driscoll2020premixed} was used to obtain the laminar flame thickness ($\delta_\mathrm{L}$) of the examined flames. Specifically, $\delta_\mathrm{L}$ was obtained using $\delta_\mathrm{L}=(\lambda^{*}/c_{\mathrm{p}}^{*})/(\rho_\mathrm{r} S_\mathrm{L,0})$, with $\lambda^{*}$ and $c_{\mathrm{p}}^{*}$ being the thermal conductivity and heat capacity of the reactants estimated at $T=\unit[1500]{K}$ and $\rho_\mathrm{r}$ estimated at $T=\unit[300]{K}$. Changing the test condition from Flame A to Flame E, $\delta_\mathrm{L}$ varied from 0.33 to \unit[0.43]{mm}. Similar to \cite{mohammadnejad2019internal, chaib2026normalization}, the effective Lewis number of the examined flames was estimated using $Le= [Le_{\mathrm{O_2}} + ALe_{\mathrm{f}}]/(1+A)$,
where indices O$_2$ and $\mathrm{f}$ refer to oxidizer and
fuel, respectively. In the above definition, $A = 1+\beta(1/\phi-1)$, in which $\beta$ is the Zel’dovich number and was obtained using the approach discussed in \cite{chaib2026normalization, moitro2026direct}. changing the test condition from Flame A to E, $\beta$ varied from 8.20 to 9.33. The fuel Lewis number was calculated using $Le_{\mathrm{f}} = x_{\mathrm{H_2}}Le_{\mathrm{H_2}} + (1- x_{\mathrm{H_2}})Le_{\mathrm{CH_4}}$, with $Le_{\mathrm{H_2}}$ and $Le_{\mathrm{CH_4}}$ being the ratio of thermal diffusivity to molecular diffusivity (in nitrogen) of hydrogen and methane~\cite{mohammadnejad2019internal}. Following~\cite{dunstan2013scalar}, $\tau$ was obtained using the products' temperature equated to the adiabatic flame temperature of the corresponding unstretched laminar premixed flame and $T_\mathrm{r}=\unit[300]{K}$. 

A separate non-reacting flow experiment was performed to characterize the background turbulence. The RMS of the velocity fluctuations were averaged inside a thin rectangular (\unit[22.2]{mm} wide and \unit[5]{mm} high) region with its center positioned at $y=$\unit[23.5]{mm} and $x=z=0$. The averaged RMS of the velocity fluctuations along $x$, $y$, and $z-$axes were 0.35, 0.24, and \unit[0.29]{$\mathrm{m\,s^{-1}}$}, respectively, resulting in the mean non-reacting turbulent-flow kinetic energy ($\overline{k}_\mathrm{nr}$) of \unit[0.132]{$\mathrm{m^2\,s^{-2}}$}. The integral length scale ($\Lambda$) and the turbulent Reynolds number ($Re_\mathrm{t}$) were obtained following the formulations presented in~\cite{mohammadnejad2020thick}, with the former and the latter being about \unit[2.7]{mm} and 57, respectively. The Kolmogorov
length scale ($\eta$) of the non-reacting flow was obtained using $\eta=\Lambda Re_\mathrm{t}^{-3/4}$, which was \unit[130]{$\tcmu$m} and was used for the estimation of the Stokes of the seed particles discussed earlier. The Karlovitz and Damk\"{o}hler ($Da$) numbers were calculated using $Ka=(u^{\prime}/S_\mathrm{L,0})^{3/2}(\Lambda/\delta_\mathrm{L})^{-1/2}$ and $Da=(u^{\prime}/S_\mathrm{L,0})^{-1}(\Lambda/\delta_\mathrm{L})$, respectively. An approach identical to that in \cite{howarth2022empirical, moitro2026direct} was followed to obtain the thermo-diffusive instability parameter ($\omega_2$), with the values listed in the last column of Table~\ref{tab: Test conditions}. As can be seen, $\omega_2$ is positive for Flames D and E, indicating these flames are thermo-diffusively unstable.    

\begin{table} \footnotesize

\caption{Experimental test conditions.} 
\addvspace{5 pt}
\scalebox{0.95}{
\centerline{\begin{tabular}{c c c c c c c c c}		
              \hline						
			&$x_{\mathrm{H_2}}$(\%) & $\phi$ & $Le$ & $\tau$ & $Ka$ &  $Da$ & $\omega_2$\\
              \hline
			 A & 0 & 0.70  & 1.03 & 5.14 & 0.84 & 4.54 & -14.3\\
             B & 20 & 0.66  & 0.98 & 4.94 & 0.86 & 4.41 & -5.4 \\  C & 40 & 0.63  & 0.90 & 4.55 & 0.88 & 4.17 & -0.9\\ 
             D & 60 & 0.57 & 0.81 & 4.38 & 0.92 & 3.84 & 2.4\\
             E &  80 &  0.51  & 0.71 & 4.01 & 0.96 & 3.49 & 4.5\\
        \hline
			\end{tabular}
		}}
		
		\label{tab: Test conditions}

\end {table}

\section{Data Reduction}
\begin{figure*}[!h]
\centerline{\includegraphics[width=0.8\textwidth]{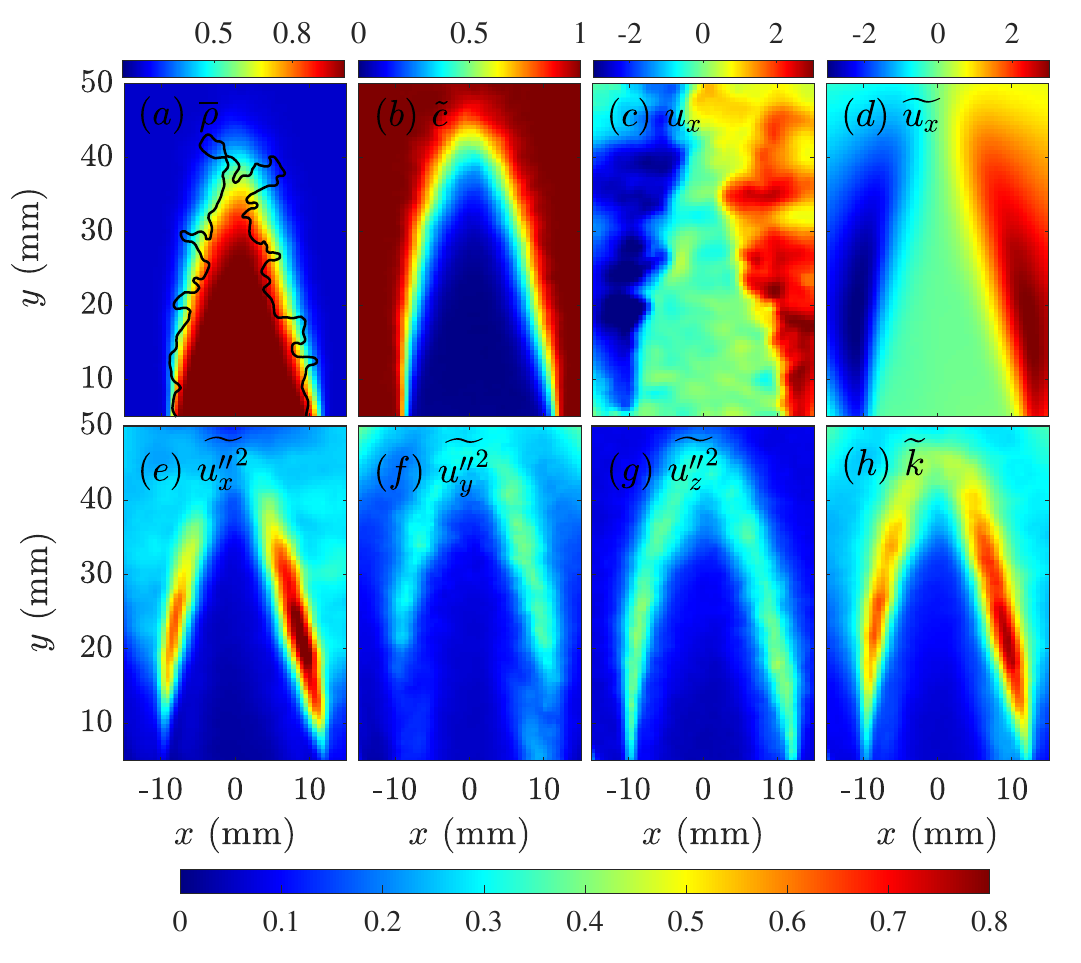}}
	\caption{(a) and (b) are the time-averaged density (in $\unit[]{kg\,m^{-3}}$) and Favre-averaged progress variable fields, respectively. The black curve overlaid in (a) is a sample instantaneous flame front. (c) and (d) are the instantaneous and Favre-averaged velocity fields (in $\unit[]{m\,s^{-1}}$) along the $x-$axis, respectively. (e)--(g) are $\widetilde{{u''_{x}}^2}$, $\widetilde{{u''_{y}}^2}$, and $\widetilde{{u''_{z}}^2}$ (in $\unit[]{m^2\,s^{-2}}$), respectively. (h) is the Favre-averaged turbulent kinetic energy field (in $\unit[]{m^2\,s^{-2}}$). The color bar at the bottom of the figure corresponds to (e--h). The results are for Flame E ($Le=0.71$).} 
   
	\label{Fig:DataReduction (TKE)}
     \vspace{-1 mm}
\end{figure*}

\vspace{-1 mm}
Figure~\ref{Fig:DataReduction (TKE)} presents the procedures to calculate $\tilde{c}$ and $\tilde{k}$ using the PIV images. The results in Fig.~\ref{Fig:DataReduction (TKE)} are for Flame E, which is chosen as a sample test condition. First, following \cite{rostami2026turbulent}, the dewarped raw PIV images were blurred using a Gaussian filter, and the adaptive thresholding was used to binarize these blurred images. Subsequently, the flame fronts were detected, with a sample flame front shown in Fig.~\ref{Fig:DataReduction (TKE)}(a), using the black curve. Then, the PIV interrogation windows (area of $a_\mathrm{w}$) were partitioned into reactants and products, with the corresponding areas of $a_\mathrm{r}$ and $a_\mathrm{p}=a_\mathrm{w}-a_\mathrm{r}$, respectively. Next, identical to \cite{altenburg2025flame, rostami2026turbulent}, the instantaneous mixture density was calculated using an area-weighted average of the reactant and product densities, i.e., $\rho = (a_\mathrm{r}/a_\mathrm{w})\rho_\mathrm{r}+(a_\mathrm{p}/a_\mathrm{w})\rho_\mathrm{p}$. In this calculation, fresh reactant and product densities were obtained using the Cantera simulations (with GRI-Mech 3.0) of freely propagating flames. The instantaneous progress variable within each integration window was also obtained using $c=a_\mathrm{p}/a_\mathrm{w}$. The instantaneous density and progress variable fields were then averaged to obtain $\overline{\rho}$ and $\tilde{c}$, with these presented in Figs.~\ref{Fig:DataReduction (TKE)}(a) and (b), respectively. These parameters ($\overline{\rho}$ and $\tilde{c}$) were additionally obtained using the Rayleigh scattering technique, and deviations of less than 25\% were observed due to the binarization process of the CPIV technique and other sources of uncertainty, such as error in the detection of the flame front. The details of the Rayleigh scattering technique and the above validation can be found in our previous study~\cite{rostami2026turbulent}. Using the instantaneous velocity and density fields, $\widetilde{u_{x}}$, $\widetilde{u_{y}}$, and $\widetilde{u_{z}}$ were obtained, with the results for a sample $u_x$ and $\widetilde{u_x}$ presented in Figs.~\ref{Fig:DataReduction (TKE)}(c) and (d), respectively. This was followed by obtaining $\widetilde{{u''_{x}}^2}$, $\widetilde{{u''_{y}}^2}$, $\widetilde{{u''_{z}}^2}$, and $\tilde{k}$, with the corresponding results presented in Figs.~\ref{Fig:DataReduction (TKE)}(e)--(h), respectively. As can be seen, the turbulence within the flame brush is anisotropic, with $\widetilde{{u''_{x}}^2}$ being larger than $\widetilde{{u''_{y}}^2}$ and $\widetilde{{u''_{z}}^2}$.

\section {Results}

The results are grouped into two subsections: (i) turbulent kinetic energy and (ii) transport of the turbulent kinetic energy, which are presented below.

\subsection{Turbulent kinetic energy}

\begin{figure}[!t]
\centerline{\includegraphics[width=0.43\textwidth]{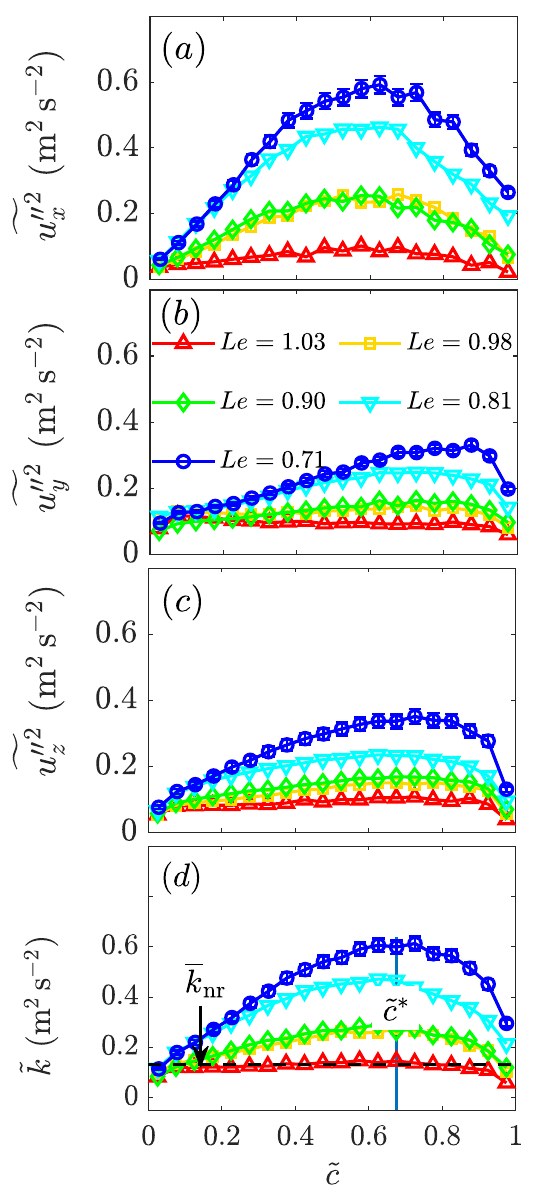}}
	\caption{(a)--(c) are the variations of $\widetilde{{u''_{x}}^2}$, $\widetilde{{u''_{y}}^2}$, and $\widetilde{{u''_{z}}^2}$ versus the Favre-averaged progress variable, respectively. (d) is the variation of the Favre-averaged turbulent kinetic energy versus the Favre-averaged combustion progress variable. }
	\label{Fig:TKEversusC}
    
  \end{figure}

This subsection focuses on the variations of several terms in Eq.~(\ref{Eq:Bray}), namely, $\tilde{k}$, $k_\mathrm{r}$, $k_\mathrm{p}$, $k_\mathrm{t}$, and $k_\mathrm{int}$, across the flame brush of the examined flames. For all test conditions, following the procedure discussed in the previous section, the variations of $\widetilde{{u''_{x}}^2}$, $\widetilde{{u''_{y}}^2}$, $\widetilde{{u''_{z}}^2}$, and $\tilde{k}$ versus $\tilde{c}$ were obtained and presented in Figs.~\ref{Fig:TKEversusC}(a)--(d), respectively. The dashed-line overlaid in Fig.~\ref{Fig:TKEversusC}(d) is $\overline{k_{\mathrm{nr}}}$. The error bars in Figs.~\ref{Fig:TKEversusC}(a)--(d) reflect the measurement uncertainty, which was quantified using a detailed analysis. For this analysis, errors in measuring the instantaneous velocity field, detecting the flame front, and binarization of the progress variable as well as density fields were propagated to quantify the combined uncertainty in measuring $\widetilde{{u''_{x}}^2}$, $\widetilde{{u''_{y}}^2}$, $\widetilde{{u''_{z}}^2}$, and $\tilde{k}$. To quantify the uncertainty due to the velocity measurements, a Monte Carlo analysis was performed. To this end, the instantaneous velocity data were perturbed using a spatio-temporally varying Gaussian noise. The standard deviation of this noise was obtained using the correlation statistics method. For each test condition, the uncertainty in the reported quantities was obtained using 1000 Monte Carlo realizations. The results showed that the maximum relative uncertainty in measuring  $\widetilde{{u''_{x}}^2}$, $\widetilde{{u''_{y}}^2}$, $\widetilde{{u''_{z}}^2}$, and $\tilde{k}$ were about 1.5\%, 1.5\%, 5.0\% and 3.0\%, respectively. To quantify the uncertainty due to flame front detection, the standard deviation of a Gaussian filter ($\sigma$) used in the flame front detection process varied between 12 and 16 pixels, and $\widetilde{{u''_{x}}^2}$, $\widetilde{{u''_{y}}^2}$, $\widetilde{{u''_{z}}^2}$, and $\tilde{k}$ were obtained for these values of $\sigma$. The results showed that the maximum relative uncertainties (due to varying $\sigma$) in measuring $\widetilde{{u''_{x}}^2}$, $\widetilde{{u''_{y}}^2}$, $\widetilde{{u''_{z}}^2}$, and $\tilde{k}$ were about 2.3\%, 2.1\%, 2.3\%, and 2.1\%, respectively. To quantify the uncertainty due to binarization of $c$ and $\rho$, similar to \cite{pfadler2008systematic, rostami2026turbulent}, the binarized $c$ and $\rho$ fields were artificially thickened using the thermal thickness of the unstretched laminar flames of the examined flames. Then, the parameters of interest were obtained for these artificially thickened flames and were compared with those calculated using the binarized flames. The comparison showed that the maximum relative uncertainties in the reported $\widetilde{{u''_{x}}^2}$, $\widetilde{{u''_{y}}^2}$, $\widetilde{{u''_{z}}^2}$, and $\tilde{k}$ were 3.6\%, 2.4\%, 3.1\%, and 2.6\%, respectively. Finally, the combined maximum relative uncertainties due to the above sources of error for measuring $\widetilde{{u''_{x}}^2}$, $\widetilde{{u''_{y}}^2}$, $\widetilde{{u''_{z}}^2}$, and $\tilde{k}$ were 4.5\%, 3.6\%,6.3\%, and 4.5\%, respectively.

Comparing the results in Figs.~\ref{Fig:TKEversusC}(a)--(c) shows that $\widetilde{{u''_{x}}^2}$ is substantially larger than $\widetilde{{u''_{y}}^2}$ and $\widetilde{{u''_{z}}^2}$, highlighting significant turbulence generation in the $x$-direction, which is nearly normal to the mean flame front at the flame base, but deviates from the normal closer to the tip of the flame brush. As can be seen in Fig.~\ref{Fig:TKEversusC}(d), except for the test condition with $Le=1.03$, $\tilde{k}$ is always larger than $\overline{k_{\mathrm{nr}}}$ and features a peak value at about $\tilde{c}^{*}=0.6-0.7$. Specifically, the ratio of the maximum value of $\tilde{k}$ to $\overline{k_\mathrm{nr}}$ increases from about 1 to 5 with decreasing $Le$ from 1.03 to 0.71. Such augmentation of the turbulent kinetic energy within the flame brush of small-Lewis-number flames is consistent with the results of past DNS studies of statistically planar~\cite{chakraborty2011effects} and jet flames~\cite{macart2018effects}.

\begin{figure}[!h]
\centerline{\includegraphics[width=0.41\textwidth]{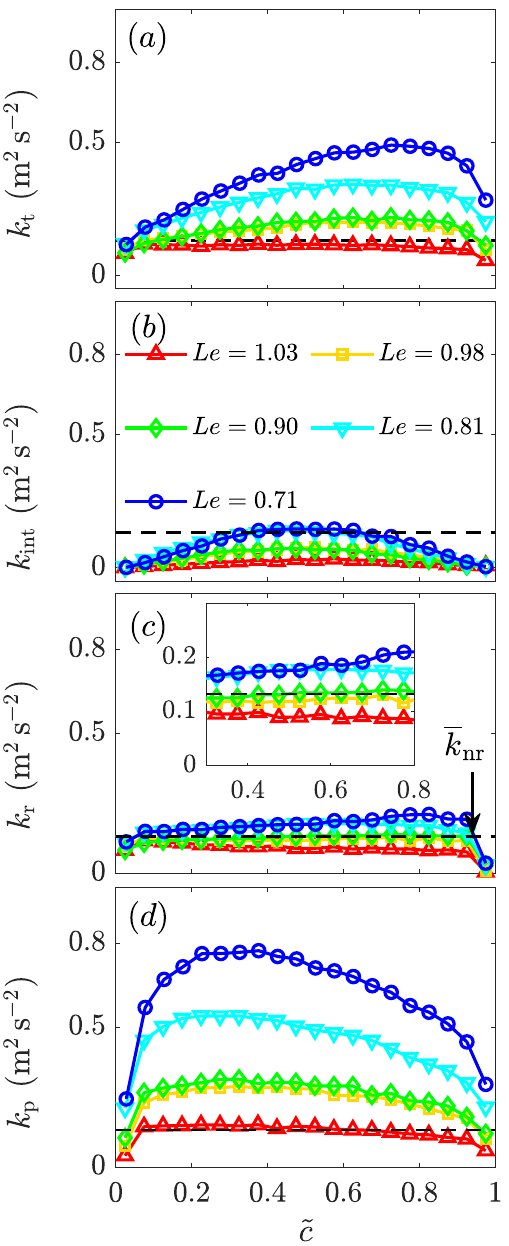}}
	\caption{(a) and (b) are the variations of the true and intermittent turbulent kinetic energy versus the Favre-averaged combustion progress variable, respectively.  (c) and (d) are the variations of the turbulent kinetic energy of the reactants and products versus the Favre-averaged combustion progress variable, respectively. The dashed lines in (a--d) are $\overline{k_\mathrm{nr}}$.} 
	\label{Fig:TrueTKEversusC}
       
\end{figure}

Investigating the contribution of the true turbulent velocity fluctuations to $\tilde{k}$, both $k_\mathrm{t}$ and $k_\mathrm{int}$ were obtained, and their variations versus $\tilde{c}$ were shown in Figs.~\ref{Fig:TrueTKEversusC}(a) and (b), respectively. As can be seen, while for $Le=1.03$, $k_\mathrm{t}$ slightly decreases within the flame brush, for flames with $Le<1$, $k_\mathrm{t}$ reaches a peak value at about $\tilde{c}^*=0.7-0.8$. Compared to $k_\mathrm{t}$, $k_{\mathrm{int}}$ features a maximum value at $\tilde{c}^* = 0.5$ as seen in Fig.~\ref{Fig:TKEversusC}(b). Nonetheless, comparison of the results in Fig.~\ref{Fig:TKEversusC}(d), Fig.~\ref{Fig:TrueTKEversusC}(a), and Fig.~\ref{Fig:TrueTKEversusC}(b) shows that $k_\mathrm{int}$ is small compared to $\tilde{k}$, suggesting that the variations of $\tilde{k}$ within the flame brush is mainly influenced by $k_\mathrm{t}$.

The variation of $k_\mathrm{t}$ across the flame brush does not allow for understanding the individual variations of the turbulent kinetic energies of the reactants ($k_\mathrm{r}$) and products ($k_\mathrm{p}$), since $k_\mathrm{t}=(1-\tilde{c})k_\mathrm{r}+\tilde{c}k_\mathrm{p}$.
To investigate this, $k_\mathrm{r}$ and $k_\mathrm{p}$ were obtained and their variations versus $\tilde{c}$ were presented in Figs.~\ref{Fig:TrueTKEversusC}(c) and (d), respectively. As can be seen, $k_\mathrm{r}$ is relatively small compared to $k_\mathrm{p}$, highlighting that the variation of $k_\mathrm{t}$ within the flame brush is primarily influenced by $k_\mathrm{p}$. It is observed that, for all examined flames and for $0.3 \lesssim  \tilde{c}< 1$,  $k_\mathrm{p}$ decreases with increasing $\tilde{c}$, with this trend being consistent with that reported in~\cite{driscoll2020premixed}. It is speculated such decreasing trend is due to the enhanced molecular mixing in the products (due to a surge in viscosity), which leads to the suppression of velocity fluctuations and, as a result, the dissipation of the products' turbulent kinetic energy. As shown in the inset of Fig.~\ref{Fig:TrueTKEversusC}(c), for flames with $Le \leq 0.81$, $k_\mathrm{r}$ slightly increases within most of the flame brush. The slight increase of $k_\mathrm{r}$ is consistent with the results of Ref.~\cite{driscoll1988measurement}. It is speculated such an increasing trend for our thermo-diffusively unstable flames (that are Flames D and E) is due to the Biot-Savart mechanism in which vorticity generated in the products (possibly due to the enhanced Baroclinic torque~\cite{chakraborty2016effects}) induces upstream velocity fluctuations in the reactants and near the flame front; hence, the increasing trend of $k_\mathrm{r}$ with $\widetilde{c}$.

\begin{figure*}[!h]
	\centerline{\includegraphics[width=0.86\textwidth]{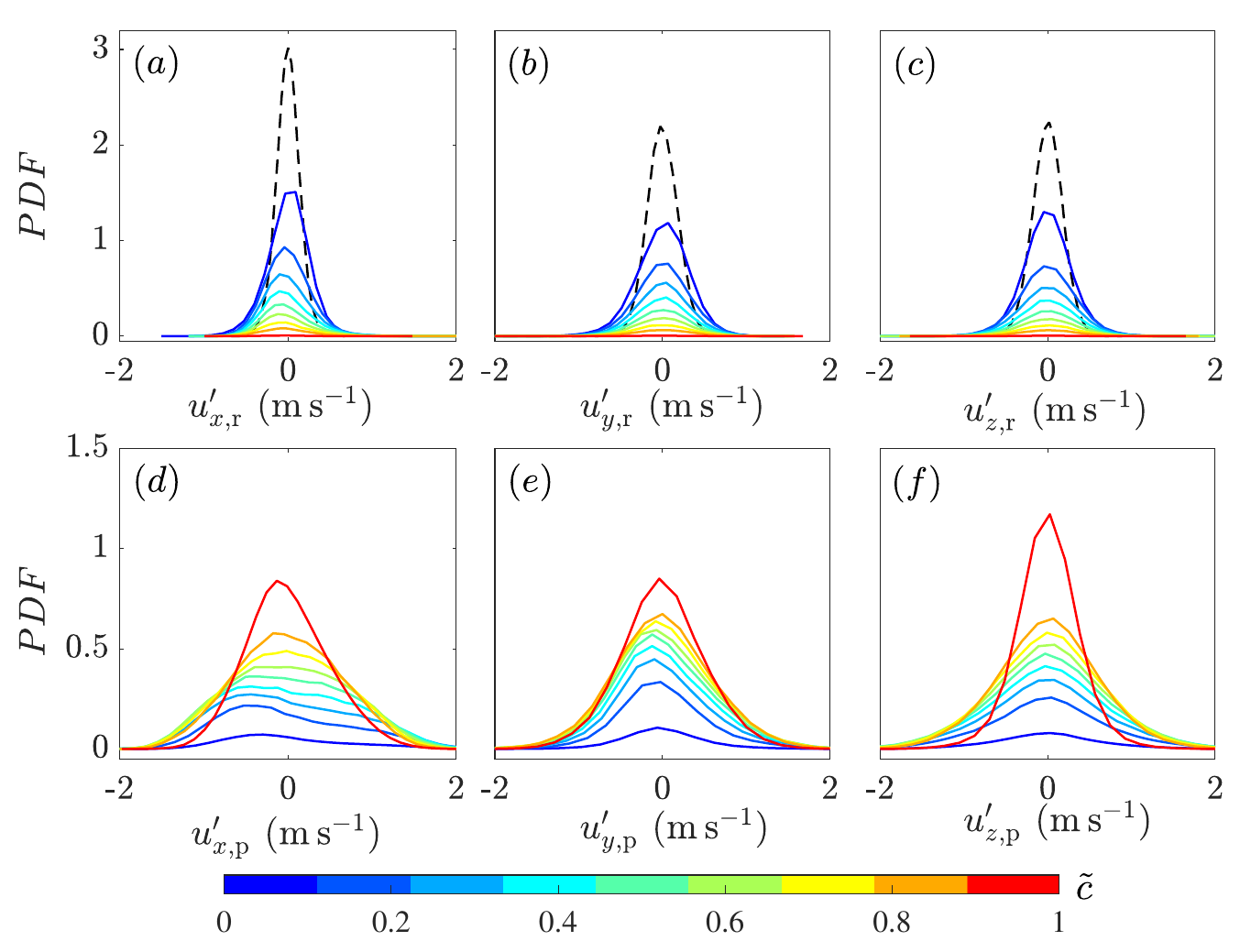}}
	\caption{(a)--(c) are the PDFs of reactants velocity fluctuations along the $x$-, $y$-, and $z$-directions, respectively. The black dashed curve is for the non-reacting condition. (d)--(f) are those in (a)--(c) for the products. The results pertain to Flame E ($Le=0.71$). } 
	\label{Fig:PDFVelocity}
      
  \end{figure*}

The above discussion shows that the flame-turbulence interaction modifies the velocity statistics of both reactants and products. For near-unity Lewis number mixtures, Driscoll and Gullati~\cite{driscoll1988measurement}  showed that such interaction not only changes the RMS of the velocity fluctuations, but also modifies the shape of the Probability Density Function (PDF) of the conditioned velocity. Specifically, the results reported in \cite{driscoll1988measurement} showed that while the PDF of reactants' velocity remained Gaussian, the PDF of the products' velocity featured a non-Gaussian distribution, contradicting the assumption used in the Bray–Moss–Libby model~\cite{bray1980turbulent}. It is of interest to investigate if such behavior holds for the non-unity effective Lewis number flames of the present study. To this end, the PDFs of the reactants and products velocity fluctuations were obtained and conditioned based on $\tilde{c}$, with the results presented in the first and second rows of Fig.~\ref{Fig:PDFVelocity}, respectively. The results are shown for Flame E ($Le = 0.71$). The results in the first to third columns of Fig.~\ref{Fig:PDFVelocity} are associated with velocity fluctuations along $x$-, $y$-, and $z$-directions, respectively. The black dashed curves overlaid on Figs.~\ref{Fig:PDFVelocity} (a)--(c) are the PDFs of $u'_{x}$, $u'_{y}$, and $u'_{z}$ for the non-reacting flow condition. To evaluate the deviation of the above PDFs from a Gaussian distribution, similar to \cite{morsy2021numerical}, the third and fourth moments of the velocity fluctuations, referred to as the skewness factor ($S$) and kurtosis ($K$), were obtained. $S$ and $K$ are 0 and 3 for a Gaussian distribution, respectively. For our non-reacting condition, the values of $S$ and $K$ were -0.04 and 3.2, respectively, confirming a nearly Gaussian distribution of the non-reacting velocity fluctuations. The slight deviations of $S$ and $K$ for the non-reacting flow from those of a Gaussian distribution are mainly due to the presence of the shear layer in jet flows, as discussed in past studies~\cite{mullin2006dual}. It was obtained that with increasing $\tilde{c}$ from 0.1 to 0.9, the values of $S$ and $K$ varied from -0.2 to 0.9 and 4.5 to 5.9 for $u'_{x,\mathrm{r}}$, and varied from 0.6 to 0.2 and 3.1 to 2.4 for $u'_{x,\mathrm{p}}$. For $u'_{y,\mathrm{r}}$, $u'_{y,\mathrm{p}}$, $u'_{z,\mathrm{r}}$, and $u'_{z,\mathrm{p}}$, while the values of $S$ were almost zero across the flame brush, the values of $K$ varied from 3.1 to 5.9. Such values of $S$ and $K$ were similar across the examined flames. Our reported non-Gaussian distribution of the PDF of the products velocity fluctuations extend that in \cite{driscoll1988measurement} to small effective Lewis number flames.

\subsection{Transport of the turbulent kinetic energy}

\begin{figure}[!h]
	\centerline{\includegraphics[width=0.41\textwidth]{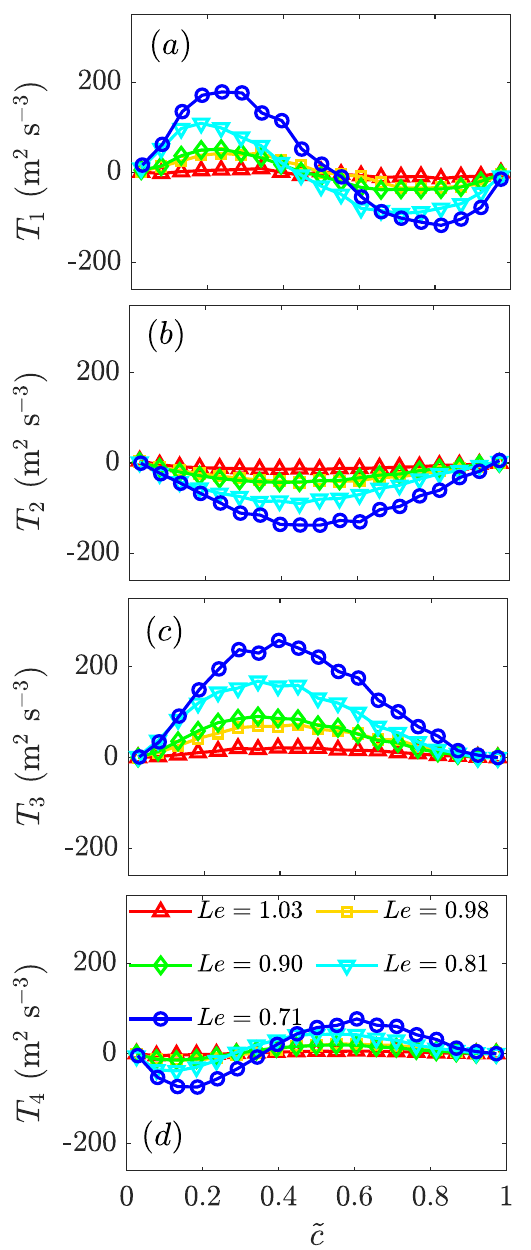}}
	\caption{(a)--(d) are the variations of $T_{1-4}$ in Eq.~(\ref{Eq.TKETransport}) versus the Favre-averaged progress variable, respectively.} 
	\label{Fig:TKEtrasnport}
    
\end{figure}

To investigate the augmentation of $\tilde{k}$ across the flame brush, $T_{1-4}$ in Eq.~(\ref{Eq.TKETransport}) were evaluated, with the results presented in Figs.~\ref{Fig:TKEtrasnport}(a)--(d), respectively. It is acknowledged that evaluating $T_{5}$ and $T_{6}$ was not possible due to the limitations of our diagnostic. Specifically, calculation of $T_{5}$ would require direct measurement of the pressure fluctuations ($p'$) within the flame brush, which was not possible in the present study. Also, calculation of $T_{6}$ would require obtaining $\mathcal{T}_{ji}= \mu[{\partial u_j}/{\partial x_i}+{\partial u_i}{\partial x_j} -(2/3)\delta_{ji}{\partial u_m}/{\partial x_m}
]$, with $\mu$ being the mixture viscosity. Key to measuring $\mathcal{T}_{ji}$ is evaluating the instantaneous dilatation rate in 3-D (i.e., measuring ${\partial u_m}/{\partial x_m}$). However, our planar measurements did not allow for obtaining the variation of $u_z$ in the out-of-plane direction, limiting our analysis to $T_{1-4}$. It is also acknowledged that, in the calculations of $T_{1-4}$, the gradient of the Favre- and time-averaged parameters in the $z$-direction were assumed to be negligible. This assumption was supported by small mean velocity in the $z$-direction of the examined small Karlovitz number flames. Indeed, the mean velocity in the $z$-direction was less than 2\% of the mean bulk flow velocity, suggesting negligible transport in this direction.

\begin{figure}[!h]
	\centerline{\includegraphics[width=0.45\textwidth]{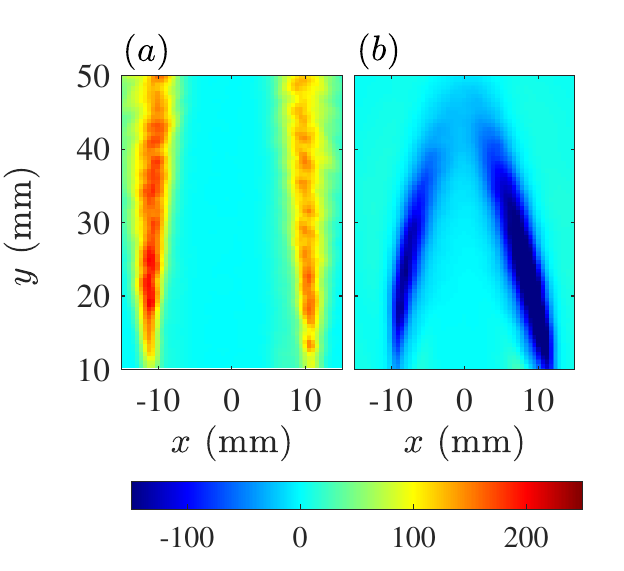}}
	\caption{(a) and (b) are the spatial distributions of the $T_{2}$ (in $\unit[]{m^2\,s^{-3}}$) for the non-reacting jet flow and flame with $Le=0.71$, respectively.} 
	\label{Fig:T2Spatial}
      
\end{figure}

The results in Fig.~\ref{Fig:TKEtrasnport}(a) show that $T_{1}$ (convection term) is positive for $\tilde{c} \lesssim 0.5$ and becomes negative for $\tilde{c}>0.5$. This trend can be explained using the variation of $\tilde{k}$, see Fig.~\ref{Fig:TKEversusC}(d), which suggested this parameter increases and then decreases with increasing $\tilde{c}$. Since the convection term is proportional to the gradient of $\tilde{k}$, $T_1$ is positive for $\tilde{c} \lesssim 0.5$ but becomes negative for $\tilde{c} > 0.5$. The results in Fig.~\ref{Fig:TKEtrasnport}(b) show that $T_{2}$ is negative within the flame brush and serves as a sink of the turbulent kinetic energy. For comparison purposes, the spatial distributions of $T_{2}$ for the non-reacting jet flow and for Flame E ($Le=0.71$) were also obtained and presented in Figs.~\ref{Fig:T2Spatial}(a) and (b), respectively. As can be seen in Fig.~\ref{Fig:T2Spatial}(a), $T_{2}$ is positive and acts as a source of the turbulent kinetic energy in the non-reacting turbulent jet flow. To elaborate the above observation, $T_{2}$ was approximated using the planar measurements of the present study and was decomposed into three constituent components ($T_{2,1-3}$) related to the streamwise normal stress, transverse normal stress, and shear stress which are given by
\begin{equation}
T_{2}
\approx
\underbrace{-\overline{\rho u_x''^{\,2}}
\frac{\partial \tilde{u}_x}{\partial x}}_{T_{2,1}}
\underbrace{-\overline{\rho u_y''^{\,2}}
\frac{\partial \tilde{u}_y}{\partial y}}_{T_{2,2}}
\underbrace{-\overline{\rho u_x''u_y''}
\left(
\frac{\partial \tilde{u}_x}{\partial y}
+
\frac{\partial \tilde{u}_y}{\partial x}
\right)}_{T_{2,3}}.
\label{Eq:T2}
\end{equation}
For non-reacting jet flows, $T_{2,1}+T_{2,2}$ was evaluated using our measurements and was shown to be negligible (compared to $T_{2,3}$), agreeing with the findings of~\cite{panchapakesan1993turbulence}. Our analysis showed that $T_{2,3}>0$, since $-\overline{\rho u_x''u_y''}<0$ and $\partial \tilde{u}_x/\partial y
+\partial\tilde{u}_y/\partial x<0$, agreeing with \cite{panchapakesan1993turbulence}. Thus, turbulence is generated due to the mean velocity gradient in the non-reacting jet flow of the present study. However, for the examined turbulent premixed flames with $Le<1$, $T_{2,1}$ was dominant and negative, due to the flow dilatation caused by heat release along the $x$-direction ($\partial \tilde{u}_x/\partial x>0$). Indeed, decreasing the effective Lewis number (at a fixed unstretched laminar flame speed) enhances the flow dilatation and increases $\overline{\rho u_x''^{\,2}}$, further increasing the absolute value of $T_2$. 

\begin{figure}[!h]
	\centerline{\includegraphics[width=0.41\textwidth]{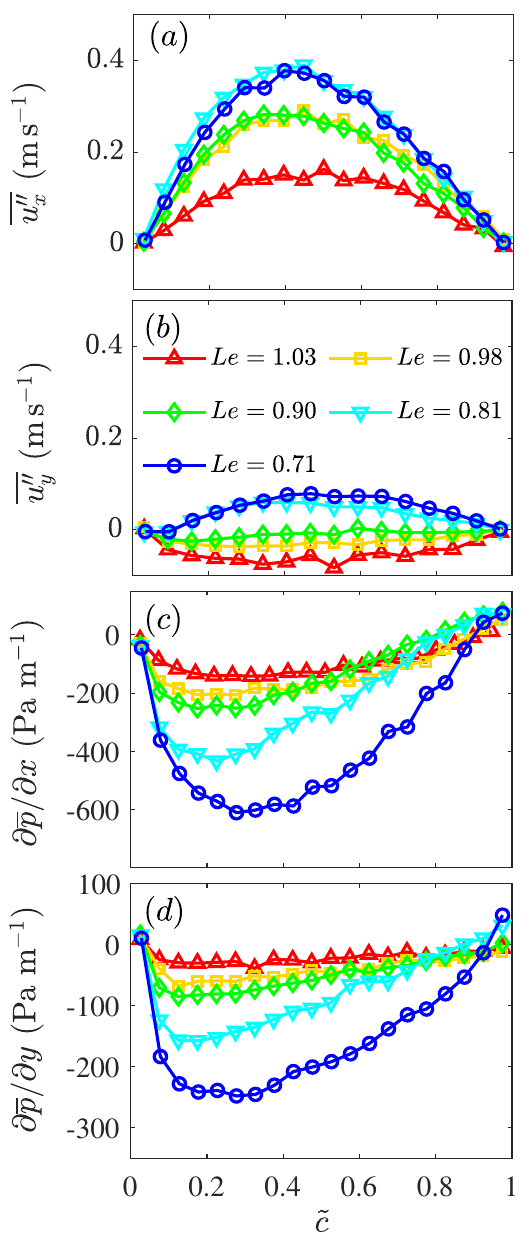}}
	\caption{(a) and (b) are the variations of the Reynolds-averaged velocity fluctuations (with respect to their Favre-averaged values) along the $x$- and $y$-directions versus the Favre-averaged combustion progress variable, respectively. (c) and (d) are the variations of the gradient of mean pressure along $x$- and $y$-directions, respectively. } 
	\label{Fig:T3components}
    
\end{figure}

\begin{figure}[!h]
	\centerline{\includegraphics[width=0.46\textwidth]{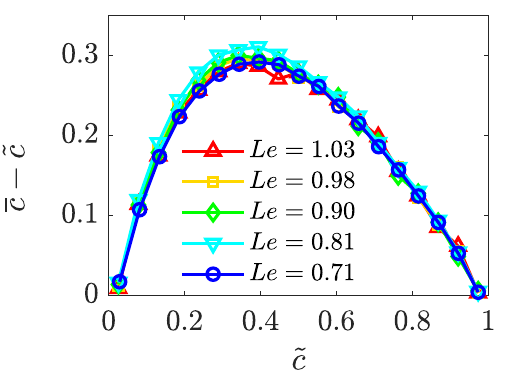}}
	\caption{Variations of the difference between Reynolds-averaged and Favre-averaged progress variables versus the Favre-averaged combustion progress variable. } 
	\label{Fig:cppbar}
 
\end{figure}

The results in Fig.~\ref{Fig:TKEtrasnport}(c) show that $T_{3}$ is positive within the flame brush, contributing to the production of $\tilde{k}$. It is also observed that the values of $T_{3}$ increase with decreasing $Le$. To elaborate this observation, the variations of contributing terms in $T_{3}$, that are $\overline{u''_{x}}$, $\overline{u''_{y}}$, $\partial\overline p/\partial x$, and $\partial\overline p/\partial y$, versus $\tilde{c}$ were obtained and presented in Figs.~\ref{Fig:T3components}(a)--(d), respectively. The results in Figs.~\ref{Fig:T3components}(c) and (d) are similar to those in our previous study~\cite{rostami2026turbulent}, except those shown here are not normalized. Comparison of the results in Figs.~\ref{Fig:T3components}(a)--(d) shows that $-\overline{u_x''}{\partial \overline{p}}/{\partial x}$ is significantly larger than $-\overline{u_y''}{\partial \overline{p}}/{\partial y}$, and as a result the former term substantially contributes to $T_3$. As can be seen in Fig.~\ref{Fig:T3components}(a), $\overline{u''_{x}}$ is positive and decreasing $Le$ increases $\overline{u''_{x}}$. Using the thin flame assumption, with its validity assessed for the examined flames of the present study in \cite{rostami2026turbulent}, it can be shown~\cite{kamal2015favre}
\begin{equation}
\label{eq:u''x}
\overline{u''_{x}}=(\overline{c}-\tilde{c})(\overline{u_{x\mathrm{,p}}}-\overline{u_{x,\mathrm{r}}}).   
\end{equation}
Evaluating the influence of $Le$ on $\overline{u''_x}$, the variations of $\overline{c}-\tilde{c}$ and $\overline{u_{x\mathrm{,p}}}-\overline{u_{x,\mathrm{r}}}$ versus $\tilde{c}$ should be examined. The latter is the slip velocity, which was shown to increase with decreasing $Le$ in~\cite{rostami2026turbulent}. The variation of the former versus $\tilde{c}$ is shown in \mbox{Fig.~\ref{Fig:cppbar}}. As can be seen, $\overline{c}-\tilde{c}$ features a peak value at $\tilde{c}=0.3-0.4$ and is almost insensitive to $Le$. Thus, it is concluded that decreasing $Le$ leads to an increase in the slip velocity, and as a result increase of $\overline{u''_{x}}$. In summary, decreasing $Le$ significantly increases both $\overline{u''_{x}}$ and $-{\partial \overline{p}}/{\partial x}$, and as a result $T_3$.

\begin{figure}[!b]
	\centerline{\includegraphics[width=0.42\textwidth]{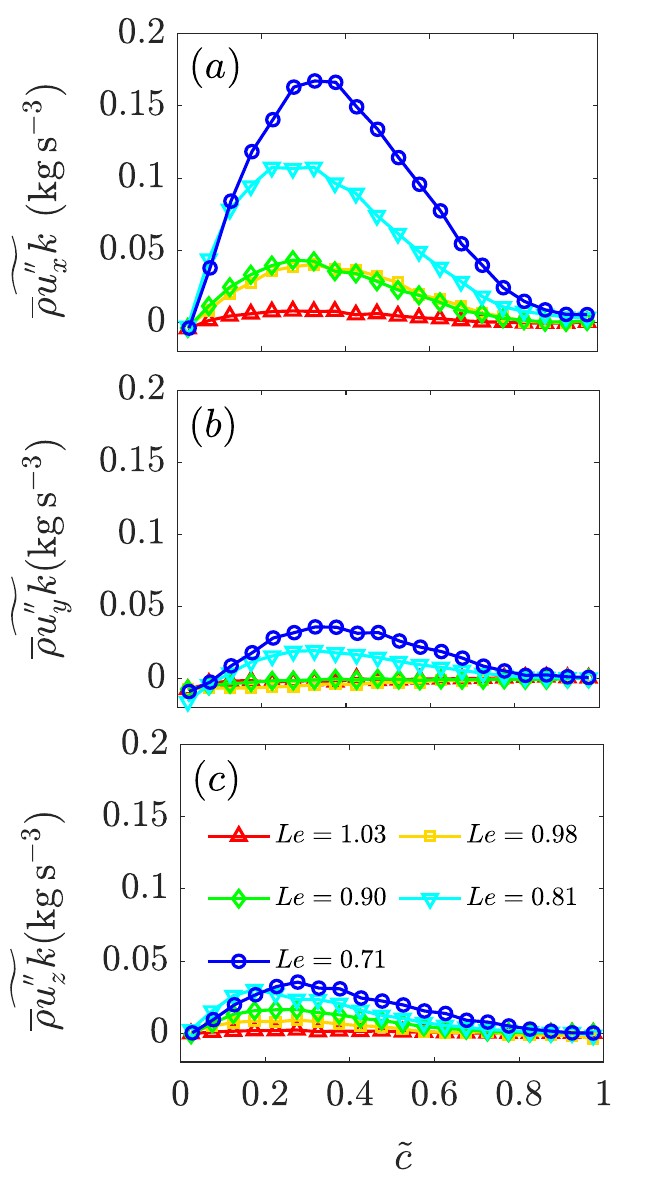}}
	\caption{(a)--(c) are the variations of the turbulent flux of the turbulent kinetic energy along $x$-, $y$-, and $z$-directions versus the Favre-averaged combustion progress variable, respectively.}   
    
	\label{Fig:TFluxK}
\end{figure}

$T_{4}$ is the negative of the divergence of the turbulent flux of the turbulent kinetic energy, see Eq.~(\ref{Eq.TKETransport}). In our previous study \cite{rostami2026turbulent}, for the flames examined here, we showed that the turbulent flux of the combustion progress variable features a counter-gradient behavior. In the DNS of \cite{chakraborty2011statistics}, it was shown that for flames featuring the counter-gradient transport of the progress variable, the turbulent flux of $k$  also features such a behavior. To investigate this, the variations of the turbulent flux of the turbulent kinetic energy along $x$-, $y$-, and $z$-directions versus $\tilde{c}$ were obtained and are presented in Figs.~\ref{Fig:TFluxK}(a)--(c), respectively. As can be seen, $\overline{\rho}\widetilde{u''_xk}$, $\overline{\rho}\widetilde{u''_yk}$, and $\overline{\rho}\widetilde{u''_zk}$ are positive, indicating the counter-gradient transport of $k$. It is also observed that decreasing $Le$ leads to enhanced counter-gradient transport, which is consistent with our previous observation \cite{rostami2026turbulent} for the turbulent flux of the combustion progress variable.

In summary, acknowledging the limitations of our study in measuring $T_5$ and $T_6$ and considering the measured variations of $T_{2-4}$, it is concluded that $T_{3}$ ($T_{2}$) is the major source (sink) term among the measured terms, contributing to the production (dissipation) of the turbulent kinetic energy. Our findings extend those concerning the augmentation of the turbulent kinetic energy reported in the DNS of small-Lewis-number statistically-planar-flames to laboratory-scale turbulent Bunsen flames.

\section{Conclusion}

The effects of flame on turbulent kinetic energy and its transport within the flame brush of small-Lewis-number turbulent premixed flames were studied experimentally. The Conditioned Particle Image Velocimetry (CPIV) was used to measure the turbulent kinetic energy and several constituent terms in its transport equation. Turbulent hydrogen-enriched methane-air flames, with the volume fraction of hydrogen varying from 0 to 80\%, were stabilized using a Bunsen burner. To isolate the effects of turbulence on the flame characteristics, the background turbulent flow was fixed. Furthermore, to isolate the effects of combustion chemistry on flame-turbulence interaction and to study the influence of the effective Lewis number on this interaction, the unstretched laminar flame speed of the examined flames was also kept constant.

The results showed that in the examined flames with the effective Lewis number of less than unity, the Favre-averaged turbulent kinetic energy maximized at a Favre-averaged combustion progress variable of about 0.6-0.7. Specifically, for the examined flames with the smallest effective Lewis number, the maximum turbulent kinetic energy within the flame brush was approximately five times that of the non-reacting flow, implying significant augmentation of the turbulent kinetic energy within the flame brush. The contribution of the unburned-burned intermittency to the Favre-averaged turbulent kinetic energy was quantified via studying the statistics of velocity conditioned based on the reactants and products. Our study showed that the contribution of intermittency to the Favre-averaged turbulent kinetic energy was relatively small, and the augmentation of the turbulent kinetic energy was due to an increase in true turbulent kinetic energy within the flame brush. The analysis of the conditioned velocity statistics showed that the flame-turbulence interaction leads to the deviation of velocity fluctuations' Probability Density Function from that of a Gaussian distribution for all examined Lewis numbers, with such observation agreeing with those in past investigations of unity Lewis number flames. 

To elaborate the augmentation of the Favre-averaged turbulent kinetic energy, several terms in the transport equation of the turbulent kinetic energy were examined. The results showed that, among the examined terms, the mean pressure gradient (dilatation) term served as a major source (sink) of the turbulent kinetic energy, contributing to the production (dissipation) of this parameter. The turbulent flux of the turbulent kinetic energy was positive across the flame brush, suggesting the occurrence of the counter-gradient behavior. Such behavior was shown to enhance with decreasing the effective Lewis number. The results of the present study agree with the trends reported in the past DNS of statistically-planar turbulent premixed flames and extend such trends to laboratory-scale flames. Our future study focuses on extending the findings presented here to intensely turbulent premixed flames.

\section*{CRediT authorship contribution statement} \addvspace{10pt}

{\bf Ali Rostami}: Conceptualization, data analysis, writing, and editing.  {\bf Sina Kheirkhah}: Funding acquisition, supervision, writing, and editing.

\section*{Declaration of competing interest} \addvspace{10pt}
The authors declare that they have no known competing financial interests or personal relationships that could have appeared to influence the work reported in this paper.

\section*{Acknowledgments}
The authors are grateful for financial support from FortisBC through Clean Growth Innovation Fund (GR036593) and the Natural Sciences and Engineering Research Council of Canada through (RGPIN-2025-04887). 

\FloatBarrier

\bibliographystyle{cnf-num}
\bibliography{cnf-refs}

@article{chakraborty2021influence,
  title={Influence of thermal expansion on fluid dynamics of turbulent premixed combustion and its modelling implications},
  author={Chakraborty, Nilanjan},
  journal={Flow, Turbul. Combust.},
  volume={106},
  pages={753--848},
  year={2021},
  publisher={Springer}
}

@misc{cantera,
    author = "David G. Goodwin and Harry K. Moffat and Ingmar Schoegl and Raymond L.
              Speth and Bryan W. Weber",
    title = "Cantera: An Object-oriented Software Toolkit for Chemical
             Kinetics, Thermodynamics, and Transport Processes",
    year = 2024,
    note = "Version 3.1.0",
    howpublished = "\url{https://www.cantera.org}",
    doi = {10.5281/zenodo.14455267}
}

@article{dunstan2013scalar,
	title={Scalar dissipation rate modelling for {L}arge {E}ddy {S}imulation of turbulent premixed flames},
	author={Dunstan, T D and Minamoto, Y and Chakraborty, N and Swaminathan, N},
	journal={Proc. Combust. Inst.},
	volume={34},
	pages={1193--1201},
	year={2013},
	publisher={Elsevier}
}

@article{mohammadnejad2020thick,
	title={Thick reaction zones in non-flamelet turbulent premixed combustion},
	author={Mohammadnejad, Sajjad and An, Qiang and Vena, Patrizio and Yun, Sean and Kheirkhah, Sina},
	journal={Combust. Flame},
	volume={222},
	pages={285--304},
	year={2020},
	publisher={Elsevier}
}

@article{bray1980turbulent,
	title={Turbulent flows with premixed reactants. {I}n {P. A.} {L}ibby and {F. A. W}illiams ({E}ds.)},
	author={Bray, K N C},
	journal={Turbulent Reacting Flows, Springer Verlag, Heidelburg, New York,},
	pages={115--183},
	year={1980},
	publisher={Springer}
}

@article{pfadler2008systematic,
  title={Systematic experiments on turbulent premixed {B}unsen flames including turbulent flux measurements},
  author={Pfadler, Sebastian and Leipertz, Alfred and Dinkelacker, Friedrich},
  journal={Combust. Flame},
  volume={152},
  pages={616--631},
  year={2008},
  publisher={Elsevier}
}

@article{lipatnikov2010effects,
  title={Effects of premixed flames on turbulence and turbulent scalar transport},
  author={Lipatnikov, AN and Chomiak, Jerzy},
  journal={Prog. Energy Combust. Sci.},
  volume={36},
  pages={1--102},
  year={2010},
  publisher={Elsevier}
}

@article{altenburg2025flame,
  title={Flame-induced pressure gradients in turbulent premixed natural gas-air and hydrogen-air jet flames},
  author={Altenburg, Luuk A and Klein, Sikke A and Tummers, Mark J},
  journal={Combust. Flame},
  volume={279},
  pages={114314},
  year={2025},
  publisher={Elsevier}
}

@article{rostami2024astigmatic,
  title={Astigmatic interferometric particle imaging of reacting {Jet} {A-1} sprays: Joint droplet and cluster characteristics},
  author={Rostami, Ali and Mohammadnejad, Sajjad and Li, Ri and Kheirkhah, Sina},
  journal={Proc. Combust. Inst.},
  volume={40},
  pages={105600},
  year={2024},
  publisher={Elsevier}
}

@article{mohammadnejad2019internal,
  title={Internal structure of hydrogen-enriched methane--air turbulent premixed flames: Flamelet and non-flamelet behavior},
  author={Mohammadnejad, Sajjad and Vena, Patrizio and Yun, Sean and Kheirkhah, Sina},
  journal={Combust. Flame},
  volume={208},
  pages={139--157},
  year={2019},
  publisher={Elsevier}
}

@article{driscoll2020premixed,
  title={Premixed flames subjected to extreme turbulence: Some questions and recent answers},
  author={Driscoll, James F and Chen, Jacqueline H and Skiba, Aaron W and Carter, Campbell D and Hawkes, Evatt R and Wang, Haiou},
  journal={Prog. Energy Combust. Sci.},
  volume={76},
  pages={100802},
  year={2020},
  publisher={Elsevier}
}

@article{wieneke2015piv,
  title={{PIV} uncertainty quantification from correlation statistics},
  author={Wieneke, Bernhard},
  journal={Measur. Science. Technol.},
  volume={26},
  pages={074002},
  year={2015},
  publisher={IOP Publishing}
}

@article{driscoll2008turbulent,
  title={Turbulent premixed combustion: Flamelet structure and its effect on turbulent burning velocities},
  author={Driscoll, James F},
  journal={Prog. Energy Combust. Sci.},
  volume={34},
  pages={91--134},
  year={2008},
  publisher={Elsevier}
}

@article{karlovitz1953studies,
  title={Studies on Turbulent flames: {A}. Flame Propagation Across velocity gradients {B}. turbulence Measurement in flames},
  author={Karlovitz, B and Denniston Jr, DW and Knapschaefer, DH and Wells, FE},
  journal={Proc. Combust. Inst.},
  volume={4},
  number={},
  pages={613--620},
  year={1953},
}

@article{scurlock1953propagation,
  title={Propagation of turbulent flames},
  author={Scurlock, AC and Grover, JH},
  journal={Proc. Combust. Inst.},
  volume={4},
  pages={645--658},
  year={1953},
  publisher={Elsevier}
}

@article{macart2018effects,
  title={Effects of combustion heat release on velocity and scalar statistics in turbulent premixed jet flames at low and high {K}arlovitz numbers},
  author={MacArt, Jonathan F and Grenga, Temistocle and Mueller, Michael E},
  journal={Combust. Flame},
  volume={191},
  pages={468--485},
  year={2018},
  publisher={Elsevier}
}

@article{bray1985unified,
  title={Unified modeling approach for premixed turbulent combustion—Part {I}: General formulation},
  author={Bray, Kenneth Noel Corbett and Libby, Paul A and Moss, JB},
  journal={Combust. Flame},
  volume={61},
  pages={87--102},
  year={1985},
  publisher={Elsevier}
}

@article{driscoll1988measurement,
  title={Measurement of various terms in the turbulent kinetic energy balance within a flame and comparison with theory},
  author={Driscoll, James F and Gulati, A},
  journal={Combust. Flame},
  volume={72},
  pages={131--152},
  year={1988},
  publisher={Elsevier}
}

@article{chakraborty2011effects,
  title={Effects of {L}ewis number on turbulent kinetic energy transport in premixed flames},
  author={Chakraborty, Nilanjan and Katragadda, Mohit and Cant, RS},
  journal={Phys. Fluids},
  volume={23},
  year={2011},
  publisher={AIP Publishing}
}

@article{nishiki2002modeling,
  title={Modeling of flame-generated turbulence based on direct numerical simulation databases},
  author={Nishiki, Shinnosuke and Hasegawa, Tatsuya and Borghi, Roland and Himeno, Ryutaro},
  journal={Proc. Combust. Inst.},
  volume={29},
  pages={2017--2022},
  year={2002},
  publisher={Elsevier}
}

@article{lipatnikov2014direct,
  title={A direct numerical simulation study of vorticity transformation in weakly turbulent premixed flames},
  author={Lipatnikov, AN and Nishiki, Shinnosuke and Hasegawa, Tatsuya},
  journal={Phys. Fluids},
  volume={26},
  year={2014},
  publisher={AIP Publishing}
}

@article{lee2020heat,
  title={Heat release effects on the {Reynolds} stress budgets in turbulent premixed jet flames at low and high {K}arlovitz numbers},
  author={Lee, Jinyoung and MacArt, Jonathan F and Mueller, Michael E},
  journal={Combust. Flame},
  volume={216},
  pages={1--8},
  year={2020},
  publisher={Elsevier}
}

@article{su2022suppression,
  title={Suppression of the turbulent kinetic energy and enhancement of the flame-normal {Reynolds} stress in premixed jet flames at small {L}ewis numbers},
  author={Su, Yunde and Lu, Zhen and Yang, Yue},
  journal={Combust. Flame},
  volume={246},
  pages={112461},
  year={2022},
  publisher={Elsevier}
}

@article{cheng1984conditional,
  title={Conditional sampling of turbulence intensities and {Reynolds} stress in premixed turbulent flames},
  author={Cheng, RK},
  journal={Combust. Sci. Techol.},
  volume={41},
  pages={109--142},
  year={1984},
  publisher={Taylor \& Francis}
}

@article{kazbekov2021flame,
  title={Flame- and flow-conditioned vorticity transport in premixed swirl combustion},
  author={Kazbekov, Askar and Steinberg, Adam M},
  journal={Proc. Combust. Inst.},
  volume={38},
  pages={2949--2956},
  year={2021},
  publisher={Elsevier}
}

@article{rising2021effects,
  title={The effects of turbulence and pressure gradients on vorticity transport in premixed bluff-body flames},
  author={Rising, Cal J and Morales, Anthony J and Geikie, Marissa K and Ahmed, Kareem A},
  journal={Phys. Fluids},
  volume={33},
  pages = {017106},
  year={2021},
  publisher={AIP Publishing}
}

@article{bobbitt2016vorticity,
  title={Vorticity transformation in high {K}arlovitz number premixed flames},
  author={Bobbitt, Brock and Lapointe, Simon and Blanquart, Guillaume},
  journal={Phys. Fluids},
  volume={28},
   pages = {015101},
  year={2016},
  publisher={AIP Publishing}
}

@article{bray1981turbulence,
  title={Turbulence production in premixed turbulent flames},
  author={Bray, KNC and Libby, Paul A and Masuya, Goro and Moss, JB},
  volume = {25},
  pages = {127--140},
  year={1981},
  journal={ Combust. Sci. Technol},
  publisher={Taylor \& Francis}
}

@article{steinberg2021structure,
  title={Structure and dynamics of highly turbulent premixed combustion},
  author={Steinberg, Adam M and Hamlington, Peter E and Zhao, Xinyu},
  journal={Prog. Energy Combust. Sci.},
  volume={85},
  pages={100900},
  year={2021},
  publisher={Elsevier}
}

@article{moitro2026direct,
  title={Direct numerical simulations of laboratory-scale thermodiffusively-unstable piloted {B}unsen flames: {F}lame surface area, turbulent flame speed, and flame height},
  author={Moitro, Arnab and Hunt, EF and Howarth, Thomas Luke and Chaib, Oussama and Bae, Jinhyun and Weller, Lee and Hochgreb, Simone and Aspden, AJ},
  journal={Combust. Flame},
  volume={288},
  pages={114916},
  year={2026},
  publisher={Elsevier}
}

@article{howarth2022empirical,
  title={An empirical characteristic scaling model for freely-propagating lean premixed hydrogen flames},
  author={Howarth, TL and Aspden, AJ},
  journal={Combust. Flame},
  volume={237},
  pages={111805},
  year={2022},
  publisher={Elsevier}
}

@article{rostami2026turbulent,
  title={Turbulent flux of the combustion progress variable and its transport in small {L}ewis number premixed flames: An experimental study},
  author={Rostami, Ali and Nahvi, Morteza and Kheirkhah, Sina},
  journal={Proc. Combust. Inst.},
  volume={42},
  pages={106545},
  year={2026},
  publisher={Elsevier}
}

@article{morsy2021numerical,
  title={Numerical and experimental study on turbulence statistics in a large fan-stirred combustion vessel},
  author={Morsy, ME and Yang, J},
  journal={Exp. Fluids},
  volume={62},
  number={5},
  pages={116},
  year={2021},
  publisher={Springer}
}

@article{mullin2006dual,
  title={Dual-plane stereo particle image velocimetry measurements of velocity gradient tensor fields in turbulent shear flow. {II}. {E}xperimental results},
  author={Mullin, John A and Dahm, Werner JA},
  journal={Phys. Fluids},
  volume={18},
  year={2006},
  publisher={AIP Publishing}
}

@online{GRI_Mech,
  author = {Smith, G P and Golden, D M and Frenklach, M and Moriarty, N W and Eiteneer, B and Goldenberg, M and Bowman, C T and Hanson, R K and Song, S and Gardiner Jr, W C and Lissianski, V V and Qin, Z},
  title = {{GRI-Mech 3.0} {(http://www.me.berkeley.edu/gri\_mech/)}},
  year = {},
  url = {},
  note = {}
}

@Misc{ZDV2023,
  author  = {Zhang, Yue and Dong, Wendi and Vandewalle, Laurien and Xu, Rui and Smith, Gregory and Wang, Hai},
  title   = {Foundational {Fuel} {Chemistry} {Model} {Version} 2.0 ({FFCM}-2)},
  journal = {FFCM-2 website},
  url     = "https://web.stanford.edu/group/haiwanglab/FFCM2",
  year    = {2023},
  }

@article{geikie2021turbulent,
  title={Turbulent flame-vortex dynamics of bluff-body premixed flames},
  author={Geikie, Marissa K and Rising, Cal J and Morales, Anthony J and Ahmed, Kareem A},
  journal={Combust. Flame},
  volume={223},
  pages={28--41},
  year={2021},
  publisher={Elsevier}
}

@article{chaib2026normalization,
  title={Normalization reveals the role of thermodiffusive effects in turbulent premixed hydrogen--methane flames},
  author={Chaib, Oussama and Bae, Jinhyun and Weller, Lee and Aspden, Andrew J and Hochgreb, Simone},
  journal={Combust. Flame},
  volume={286},
  pages={114852},
  year={2026},
  publisher={Elsevier}
}

@article{chakraborty2011statistics,
  title={Statistics and modelling of turbulent kinetic energy transport in different regimes of premixed combustion},
  author={Chakraborty, Nilanjan and Katragadda, Mohit and Cant, R Stewart},
  journal={Flow, Turbul. Combust.},
  volume={87},
  pages={205--235},
  year={2011},
  publisher={Springer}
}

@article{chakraborty2016effects,
  title={Effects of {L}ewis number on vorticity and enstrophy transport in turbulent premixed flames},
  author={Chakraborty, Nilanjan and Konstantinou, Ilias and Lipatnikov, Andrei},
  journal={Phys. Fluids},
  volume={28},
  year={2016},
  publisher={AIP Publishing}
}

@article{kamal2015favre,
  title={Favre-and {R}eynolds-averaged velocity measurements: Interpreting {PIV} and {LDA} measurements in combustion},
  author={Kamal, M Mustafa and Zhou, Ruigang and Balusamy, Saravanan and Hochgreb, Simone},
  journal={Proc. Combust. Inst.},
  volume={35},
  pages={3803--3811},
  year={2015},
  publisher={Elsevier}
}

@article{panchapakesan1993turbulence,
  title={Turbulence measurements in axisymmetric jets of air and helium. {P}art 1. {A}ir jet},
  author={Panchapakesan, Nagangudy R and Lumley, John L},
  journal={J. Fluid Mech.},
  volume={246},
  pages={197--223},
  year={1993},
  publisher={Cambridge University Press}
}

@article{picano2011dynamics,
  title={Dynamics of {PIV} seeding particles in turbulent premixed flames},
  author={Picano, Francesco and Battista, Francesco and Troiani, Guido and Casciola, Carlo Massimo},
  journal={Exp. Fluids},
  volume={50},
  pages={75--88},
  year={2011},
  publisher={Springer}
}

\end{document}